\documentclass[runningheads]{llncs}
\usepackage{amsmath}
\usepackage{amssymb}
\usepackage{graphicx}
\usepackage{pifont}
\newcommand{\cmark}{\ding{51}}%
\newcommand{\xmark}{\ding{55}}%
\usepackage[hidelinks]{hyperref}
\usepackage{color}

\newcommand{\R}{\mathbb{R}}
\begin{document}
\title{TabAttention: Learning Attention Conditionally on Tabular Data}
%
%
\author{Michal K. Grzeszczyk\inst{1} \and Szymon Płotka\inst{1, 2, 3} \and Beata Rebizant\inst{4} \and Katarzyna Kosińska-Kaczyńska\inst{4} \and Michał Lipa\inst{5} \and Robert Brawura-Biskupski-Samaha\inst{4} \and Przemysław Korzeniowski\inst{1} \and
Tomasz Trzciński \inst{6, 7, 8} \and
Arkadiusz Sitek \inst{9}}


\authorrunning{M. K. Grzeszczyk et al.}

\newcommand\blfootnote[1]{%
  \begingroup
  \renewcommand\thefootnote{}\footnote{#1}%
  \addtocounter{footnote}{-1}%
  \endgroup
}

\institute{Sano Centre for Computational Medicine, Cracow, Poland \email{m.grzeszczyk@sanoscience.org} \and
Informatics Institute, University of Amsterdam, Amsterdam, The Netherlands \and
Amsterdam University Medical Center, Amsterdam, The Netherlands \and
The Medical Centre of Postgraduate Education, Warsaw, Poland \and
Medical University of Warsaw, Warsaw, Poland \and
Warsaw University of Technology, Warsaw, Poland \and 
IDEAS NCBR, Warsaw, Poland \and
Tooploox, Wroclaw, Poland \and
Massachusetts General Hospital, Harvard Medical School, Boston, MA, USA}
\maketitle              
\begin{abstract}
 
Medical data analysis often combines both imaging and tabular data processing using machine learning algorithms. While previous studies have investigated the impact of attention mechanisms on deep learning models, few have explored integrating attention modules and tabular data. In this paper, we introduce TabAttention, a novel module that enhances the performance of Convolutional Neural Networks (CNNs) with an attention mechanism that is trained conditionally on tabular data. Specifically, we extend the Convolutional Block Attention Module to 3D by adding a Temporal Attention Module that uses multi-head self-attention to learn attention maps. Furthermore, we enhance all attention modules by integrating tabular data embeddings. Our approach is demonstrated on the fetal birth weight (FBW) estimation task, using 92 fetal abdominal ultrasound video scans and fetal biometry measurements. Our results indicate that TabAttention outperforms clinicians and existing methods that rely on tabular and/or imaging data for FBW prediction. This novel approach has the potential to improve computer-aided diagnosis in various clinical workflows where imaging and tabular data are combined. We provide a source code for integrating TabAttention in CNNs at \url{https://github.com/SanoScience/Tab-Attention}.

\end{abstract}

\section{Introduction}

Many clinical procedures involve collecting data samples in the form of imaging and tabular data. 
New deep learning (DL) architectures fusing image and non-image data are being developed to extract knowledge from both sources of information and improve predictive capabilities \cite{huang2020fusion}. While concatenation of tabular and imaging features in final layers is widely used \cite{holste2021end,liu2018joint}, this approach limits the interaction between them. To facilitate better knowledge transfer between these modalities more advanced techniques have been proposed. Duanmu \textit{et al.} \cite{duanmu2020prediction} presented the Interactive network in which tabular features are passed through a separate branch and channel-wise multiplied with imaging features at different stages of Convolutional Neural Network (CNN). Pölsterl \textit{et al.} \cite{polsterl2021combining} proposed a Dynamic Affine Feature Map Transform (DAFT) to shift and scale feature maps conditionally on tabular data. In \cite{guan2021predicting}, Guan \textit{et al.} presented a method for transforming tabular data and processing them together with 3D feature maps via VisText self-attention module. 
The importance of the attention mechanism on DL models' performance has been extensively studied \cite{vaswani2017attention}. Convolutional Block Attention Module (CBAM) \cite{woo2018cbam} has been shown to improve the performance of DL models on high dimensional data \cite{wang2022spatiotemporal,yadav2020frequency}. Despite these advances, few studies have explored the potential of incorporating attention maps with imaging and tabular data simultaneously.

We develop such a solution and as an example of application, we use fetal birth weight (FBW) prediction from ultrasound (US) data. It is a challenging task requiring clinicians to collect US videos of fetal body parts and fetal biometry measurements. Currently, abdominal circumference (AC), head circumference (HC), biparietal diameter (BPD), and femur length (FL) are used to estimate FBW with heuristic formulae \cite{hadlock1985estimation}. The predicted weight is the indicator of perinatal health prognosis or complications in pregnancy and has an impact on the method of delivery (vaginal or Cesarean) \cite{pressman2000prediction}. Unfortunately, the current approach to FBW estimation is often imprecise and can lead to a mean absolute percentage error (MAPE) of 10\%, even if performed by experienced sonographers \cite{sherman1998comparison}. 
An ensemble of Machine Learning algorithms was proposed by Lu \textit{et al.} \cite{lu2019ensemble} for solving this task. CNNs are applied for fetal biometry measurements estimation from US standard planes \cite{bano2021autofb} or US videos \cite{plotka2022deep}. Tao \textit{et al.} \cite{tao2021fetal} approach this problem with a recurrent network utilizing temporal features of fetal weight changes over weeks concatenated with fetal parameters. Płotka \textit{et al.} \cite{plotka2022babynet} developed BabyNet, a hybrid CNN with Transformer layers to estimate FBW directly from US videos. Recent studies show that there is a strong correlation between the image features of the abdominal plane and the estimated fetal weight, indicating that it can serve as a dependable indicator for evaluating fetal growth \cite{campbell1975ultrasonic}. We utilize the US videos of the abdomen (imaging data) and biometry measurements with other numerical values (tabular data) during our experiments. 

In this work, we introduce TabAttention, a novel module designed to enhance the performance of CNNs by incorporating tabular data. TabAttention extends the CBAM to the temporal dimension by adding a Temporal Attention Module (TAM) that leverages Multi-Head Self-Attention (MHSA) \cite{vaswani2017attention}. Our method utilizes pooled information from imaging feature maps and tabular data (represented as tabular embeddings) to generate attention maps through Channel Attention Module (CAM), Spatial Attention Module (SAM), and TAM. By incorporating tabular data, TabAttention enables the network to better identify \textit{what}, \textit{where}, and \textit{when} to focus on, thereby improving performance. We evaluate our method on the task of estimating FBW from abdominal US videos and demonstrate that TabAttention is at least on par with existing methods, including those based on tabular and/or imaging data, as well as clinicians. The main contributions of our work are: 1) the introduction of TabAttention, a module for conditional attention learning with tabular data, 2) the extension of CBAM to the temporal dimension via the TAM module, and 3) the validation of our method on the FBW estimation task, where we demonstrate that it is competitive with state-of-the-art methods.

\section{Method}

\begin{figure}[t!]
    \centering
    \includegraphics[width=\textwidth]{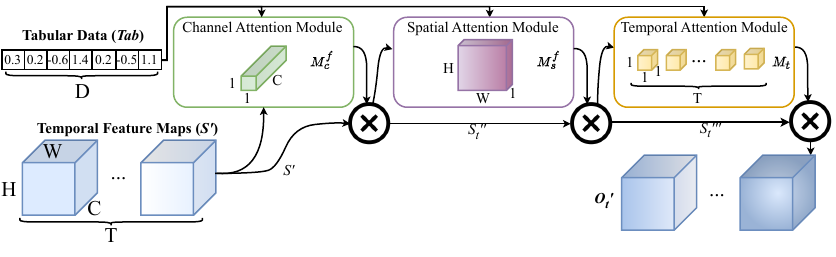}
    \caption{The overview of proposed TabAttention module inspired by CBAM \cite{woo2018cbam}. We add a Temporal Attention Module to the existing architecture to extend the method to 3D data processing. In our setting, channel, spatial and temporal attention maps are trained conditionally on tabular data. $\otimes$ indicates element-wise multiplication.}
    \label{fig:overview}
\end{figure}

In this section, we introduce the fundamental components of the TabAttention module. We detail the development of CBAM augmented with a Temporal Attention Module. Then, we elaborate on how TabAttention leverages tabular embeddings to modulate the creation of attention maps and outline how the module can be seamlessly incorporated into the residual block of ResNet.

Fig.~\ref{fig:overview} presents the overview of the TabAttention module. Given US video sequence $S \in \R^{T_{0} \times 1 \times H_{0} \times W_{0}}$ of height $H_0$, width $W_0$ and frame number $T_0$ as the input, 3D CNN produces intermediate temporal feature maps $S' \in \R^{T \times C \times H \times W}$ where $C$ is the number of channels. In our setting, the CBAM block generates $T$ 1D channel attention maps $M_{c} \in \R^{T \times C \times 1 \times 1}$ and $T$ 2D spatial attention maps $M_{s} \in \R^{T \times 1 \times H \times W}$. We create attention maps separately for every temporal feature map as the information of \textit{what} is meaningful and \textit{where} it is important to focus on might change along the temporal dimension. To account for the temporal changes and focus on \textit{when} is the informative part we add TAM which infers temporal attention map $M_{t} \in \R^{T \times 1 \times 1 \times 1}$. Intermediate temporal feature maps $S'$ are refined with attention maps in the following way:
\begin{equation}
        S''  =  M_{c}(S') \otimes S' \quad \quad \quad S''' = M_{s}(S'') \otimes S'' \quad \quad \quad \boldsymbol{O'} = M_{t}(S''') \otimes S'''
\end{equation}
\noindent Here $\boldsymbol{O'}$ denotes the output of the module and $\otimes$ is an element-wise multiplication during which attention maps are broadcasted along all unitary dimensions.

\begin{figure}[t!]
    \centering
    \includegraphics[width=\textwidth]{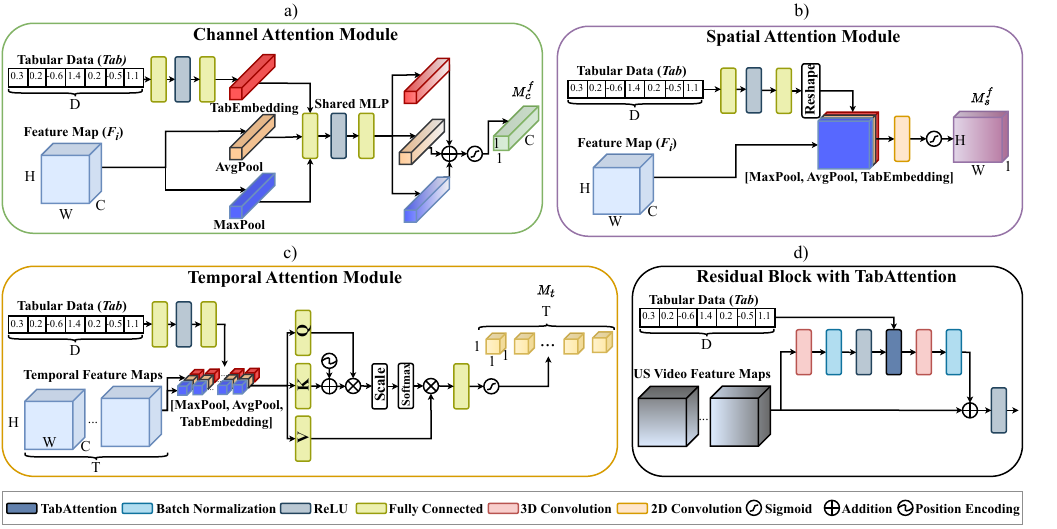}
    \caption{Details of TabAttention components: CAM (a), SAM (b) and TAM (c) with tabular embeddings. In TAM, only one attention head is visualised. TabAttention is integrated with residual block as presented in (d). $\otimes$ indicates matrix multiplication.}
    \label{fig:detailed}
\end{figure}

In general, attention maps are computed based on information aggregated by average- and max-pooling along specified dimensions which are then passed through shared layers for refinement (Fig. \ref{fig:detailed}). Then, these refined descriptors are passed through the sigmoid function to create final attention maps. To account for the tabular information during attention maps computing, we embed the input tabular data $Tab \in \R^{D}$, where $D$ is the number of numerical features, with two linear layers and Rectified Linear Unit (ReLU) activation in between. The tabular data is embedded to the size of pooled feature maps. The embedding is passed through shared layers in the same way as pooled feature maps. Therefore, the attention maps are computed conditionally on tabular data. Thus, the output of TabAttention $\boldsymbol{O'_{t}}$ is computed as follows:
\begin{equation}
       S''_{t} = M_{c}(S', Tab) \otimes S' \quad \quad S'''_{t} = M_{s}(S''_{t}, Tab) \otimes S''_{t} \quad \quad    \boldsymbol{O'_{t}} = M_{t}(S'''_{t}, Tab) \otimes S'''_{t}
\end{equation}

\noindent \textbf{Channel Attention Module.} We follow the design of the original CBAM \cite{woo2018cbam}. We split temporal feature maps into $T$ feature maps $F_{i}$ where $i\in 1, ..., T$ so that each of them is passed through CAM separately. To compute the channel attention ($M_{c}$), we aggregate the spatial information through average- and max-pooling to produce descriptors ($F^{c}_{avg_{i}}$, $F^{c}_{max_{i}} \in \R^{C \times 1 \times 1}$). We pass the tabular data through a multi-layer perceptron ($MLP_{emb_{c}}$) with one hidden layer (of size $\R^{\frac{C}{z}}$, where $z$ is the reduction ratio set to 16) and ReLU activation to embed it into the same dimension as spatial descriptors. Then, both descriptors, with tabular embedding are passed through the shared network which is MLP with a hidden activation size of $\R^{\frac{C}{z}}$ and one ReLU activation. After the MLP is applied, the output vectors are element-wise summed to produce the attention map. We concatenate attention maps of all feature maps to produce $M_{c}$:
\begin{align}
       &M_{c}(S', Tab) = [M^{f}_{c}(F_{i}, Tab)]_{i=1, ..., T}\\
       &M^{f}_{c}(F_{i}, Tab) = \sigma(MLP(F^{c}_{max_{i}}) + MLP(F^{c}_{avg_{i}}) + MLP(MLP_{emb_{c}}(Tab))) 
\end{align}
\noindent \textbf{Spatial Attention Module.} After splitting the temporal feature maps, we average- and max-pool them along channel dimension to produce feature descriptors ($F^{s}_{avg_{i}}$, $F^{s}_{max_{i}} \in \R^{1 \times H \times W}$). We pass the tabular data through $MLP_{emb_{s}}$ with one hidden layer of size $\R^{\frac{H \times W}{2}}$ and ReLU activation to embed it into the same dimension as spatial descriptors. We reshape this embedding to the size of feature descriptors and concatenate it with them. We pass the following representation through a 2D convolution layer and the sigmoid activation:
\begin{align}
       &M_{s}(S'', Tab) = [M^{f}_{s}(F_{i}, Tab)]_{i=1, ..., T}\\
       &M^{f}_{s}(F_{i}, Tab) = \sigma(Conv([F^{s}_{max_{i}}, F^{s}_{avg_{i}}, Reshape(MLP_{emb_{s}}(Tab))])) 
\end{align}
\noindent \textbf{Temporal Attention Module.} We create temporal descriptors by average- and max-pooling temporal feature maps along all non-temporal dimensions ($F^{t}_{avg_{i}}$, $F^{t}_{max_{i}} \in \R^{T \times 1 \times 1 \times 1}$). We embed tabular data with $MLP_{emb_{t}}$ with one hidden layer of size $\R^{\frac{T}{2}}$ into the same dimension. We concatenate  created vectors and treat them as the embedding of the US sequence which we pass to the MHSA layer (with 2 heads). We create the query (Q), key (K) and value (V) with linear layers and an output size of $d$ (4). We add relative positional encodings \cite{shaw2018self} $r$ to K. After passing through MHSA, we squash the refined representation with one MLP layer and sigmoid function to create a temporal attention map $M_{t}$: 

\begin{align}
       & MHSA(S_{emb}) = MLP\left(\left[softmax\left(\frac{Q_j(K_j+r)^T}{\sqrt{d}}\right)V_j\right]_{j=1, 2}\right)\\
       & M_{t}(S''', Tab) = \sigma(MHSA([F^{t}_{max}, F^{t}_{avg}, MLP_{emb_{t}}(Tab)])) 
\end{align}

\noindent
TabAttention can be integrated within any 3D CNN (or 2D CNN in case TAM is omitted). As illustrated in Fig. \ref{fig:detailed}, we add TabAttention between the first ReLU and the second convolution in the residual block to integrate our module with 3D ResNet-18.

\section{Experiments and Results}
This section describes the dataset used and provides implementation details of our proposed method. We benchmark the performance of TabAttention against several state-of-the-art methods and compare them to results obtained by clinicians. Additionally, we conduct an ablation study to demonstrate the significance of each key component utilized in our approach.

\begin{figure}[t!]
    \centering
    \includegraphics[width=\textwidth]{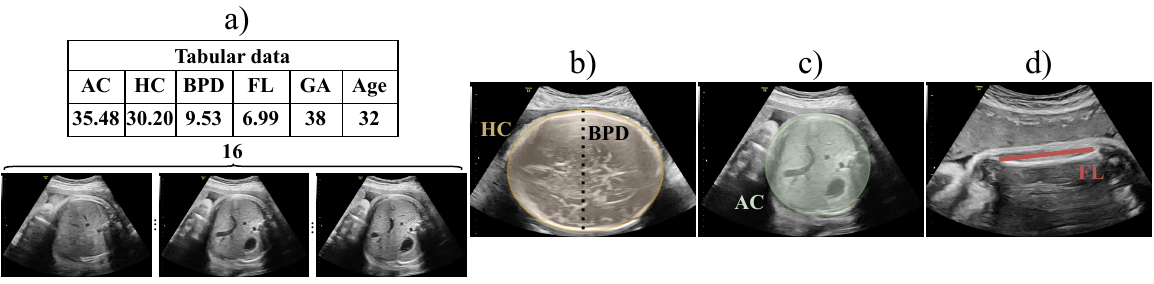}
    \caption{Exemplary input to our method (a) of tabular data and frames of abdominal scans. 
    Fetal US scans of the head (b), abdomen (c), and femur (d) were used to obtain AC, HC, BPD and FL utilized in tabular data.}
    \label{fig:data}
\end{figure}

\noindent
\textbf{Dataset.} This study was approved by the Ethics Committee of the Medical University of Warsaw (Reference KB.195/2021) and informed consent was obtained for all subjects. The multi-site dataset was acquired using international standards approved by \cite{salomon2019isuog}. The dataset consists of 92 2D fetal US video scans captured in the standard abdominal plane view. These scans were collected from 92 pregnant women (31.89 $\pm$ 4.76 years), across three medical centers, and obtained as part of routine US examination done less than 24 hours before delivery. This allowed us to obtain the real ground truth which was baby weight soon after birth. 
Five experienced sonographers (14.2 $\pm$ 4.02 years of experience) acquired the data using a single manufacturer device (General Electric) and several models (GE Voluson E6, S8, P8, E10, and S10). The abdominal fetal US videos (5-10 seconds, 13-37 frames per second) were saved in the DICOM file format. We resized the pixel spacing to 0.2 mm $\times$ 0.2 mm for all video clips. As tabular data, we used six numerical features: AC (34.51 $\pm$ 2.35 cm), HC (33.56 $\pm$ 1.41 cm), BPD (9.40 $\pm$ 0.46 cm), FL (7.28 $\pm$ 0.33 cm), GA (38.29 $\pm$ 1.47 weeks), and mother's age. The examples of how the measurements were obtained are presented in Fig.~\ref{fig:data}.  The actual birth weight of the fetus obtained right post-delivery (3495 $\pm$ 507 grams) was used as the target of the prediction.

\noindent
\textbf{Implementation details.} We use 3D ResNet-18 as our base model. We implement all experiments with PyTorch and train networks using NVIDIA A100 80GB GPU for 250 epochs with a batch size of 16 and an initial learning rate chosen with grid search from the set of $\{1 \times 10^{-2}, 1 \times 10^{-3}, 1 \times 10^{-4}\}$. To minimize the Mean Squared Error loss function, we employ the Adam \cite{kingma} optimizer with L2 regularization of $1 \times 10^{-4}$ and cosine annealing learning rate scheduler. To evaluate the reliability of the regression algorithm, we conduct five-fold cross-validation (CV) and ensure that each patient's data is present in only one fold. To ensure similar birth weight distribution in all folds, we stratify them based on the assignment of data samples into three bins: $<$ 3000 g, $>$ 4000 g, and in-between. The input frames are of size 128 $\times$ 128 pixels. We follow the approach presented in \cite{plotka2022babynet}, we set the number of input frames to 16 and average per-patient predictions of all 16 frame segments from the single video. Throughout the training process, we employ various data augmentation techniques such as rotation, random adjustments to brightness and contrast, the addition of Gaussian noise, horizontal flipping, image compression, and motion blurring for every batch. We standardize all numerical features to a mean of 0 and a standard deviation of 1. We use Root Mean Square Error (RMSE), Mean Absolute Error (MAE), and MAPE to evaluate the regression performance.

\noindent
\textbf{Comparison with state-of-the-art methods.} We compare TabAttention with several methods utilizing tabular data only (Linear Regression \cite{scikit-learn}, XGBoost \cite{xgboost}), imaging data only (3D ResNet-18 \cite{tran2018closer}, BabyNet \cite{plotka2022babynet}), both types of data (Interactive \cite{duanmu2020prediction}, DAFT \cite{polsterl2021combining}), and Clinicians. The predictions of Clinicians were achieved using Hadlock III \cite{hadlock1985estimation} formula and AC, HC, BPD, FL measurements. The comparison of results from the five-fold CV is presented in Table \ref{tab:resultssota}. TabAttention achieves the lowest MAE, RMSE and MAPE (170 $\pm$ 26, 225 $\pm$ 37, 5.0 $\pm$ 0.8 respectively) among all tested methods. Our approach outperforms clinically utilized heuristic formulae, machine learning, and image-only DL methods (two-tailed paired t-test p-value $<$ 0.05). Results of TabAttention are also best compared with all DL models utilizing tabular and imaging modalities, however, the difference does not reach statistical significance with a p-value around 0.11.

\bgroup
\def\arraystretch{1.01}%
\begin{table}[t!]
    \caption{Five-fold cross-validation results of state-of-the-art methods utilizing imaging (Img.) and/or tabular (Tab.) data. The mean of MAE, RMSE, and MAPE across all folds are presented. The best results are bolded.}
    \label{tab:resultssota}
    \begin{center}
        \begin{tabular}{c|c|c|c|c|c}
            \textbf{Method} &\textbf{Img.} & \textbf{Tab.}& \textbf{mMAE [g]} & \textbf{mRMSE [g]} & \textbf{mMAPE [\%]}\\
            \hline
            BabyNet \cite{plotka2022babynet}        & \cmark & \xmark & $294 \pm 30$ & $386 \pm 56$ & $8.5 \pm 1.0$ \\ \hline
            3D ResNet-18 \cite{tran2018closer}      & \cmark & \xmark & $289 \pm 38$ & $373 \pm 43$ & $8.5 \pm 1.1$ \\ \hline
            XGBoost                                 & \xmark & \cmark & $259 \pm 23$ & $328 \pm 26$ & $7.6 \pm 0.1$ \\ \hline
            Linear Regression                       & \xmark & \cmark & $207 \pm 18$ & $260 \pm 19$ & $6.0 \pm 0.1$ \\ \hline
            Clinicians                              & \xmark & \cmark & $205 \pm 14$ & $253 \pm 13$ & $5.9 \pm 0.0$ \\ \hline
            DAFT \cite{polsterl2021combining}       & \cmark & \cmark & $175 \pm 30$ & $244 \pm 42$ & $5.3 \pm 1.0$ \\ \hline
            Interactive \cite{duanmu2020prediction} & \cmark & \cmark & $172 \pm 27$ & $230 \pm 44$ & $5.2 \pm 0.9$ \\ \hline
            \textbf{TabAttention (ours)}            & \cmark & \cmark & $\mathbf{170 \pm 26}$ & $\mathbf{225 \pm 37}$ & $\mathbf{5.0 \pm 0.8}$\\ 
        \end{tabular}
    \end{center}
\end{table}
\egroup

\noindent
\textbf{Ablation study.} We conduct ablation experiments to validate the effectiveness of key components of our proposed method (Table \ref{tab:ablation}). We employ 3D ResNet-18 as the baseline model. The integration of TAM or CBAM with attention maps learned conditionally on tabular data into the 3D ResNet-18 architecture improves the predictive performance of the network. Subsequently, the incorporation of full TabAttention further enhances its capabilities.

\bgroup
\def\arraystretch{1.1}%
\begin{table}[t!]
    \caption{Five-fold cross-validation results of ablation study with key components of TabAttention. The first row is the result of the baseline method. The next rows refer to modules of TabAttention with or without tabular embeddings (Tab.) and the last one is full TabAttention.}\label{tab:ablation}
    \begin{center}
        \begin{tabular}{c|c|c|c|c|c}
            \textbf{Method}&\textbf{Img.} & \textbf{Tab.} & \textbf{mMAE [g]} & \textbf{mRMSE [g]} & \textbf{mMAPE [\%]}\\
            \hline
            3D ResNet-18 \cite{tran2018closer}      & \cmark & \xmark & $289 \pm 38$ & $373 \pm 43$ & $8.5 \pm 1.1$ \\ \hline
            + TAM                                   & \cmark & \xmark & $288 \pm 43$ & $389 \pm 65$ & $8.4 \pm 1.2$ \\ \hline
            + CBAM + Tab.                           & \cmark & \cmark & $271 \pm 51$ & $371 \pm 99$ & $7.7 \pm 1.3$ \\ \hline
            + TAM + Tab.                            & \cmark & \cmark & $180 \pm 32$ & $237 \pm 45$ & $5.5 \pm 0.1$\\ \hline
            + \textbf{TabAttention (ours)}          & \cmark & \cmark & $\mathbf{170 \pm 26}$ & $\mathbf{225 \pm 37}$ & $\mathbf{5.0 \pm 0.8}$\\ 
        \end{tabular}
    \end{center}
\end{table}
\egroup

\section{Discussion and Conclusions}

In this work, we present a novel method, TabAttention, that can effectively compete with current state-of-the-art image and/or tabular-based approaches in estimating FBW. We found that it outperformed Clinicians achieving mMAPE of 5.0\% vs. 5.9\% (p-value $<$ 0.05). A key advantage of our approach is that it does not require any additional effort from clinicians since the necessary data is already collected as part of standard procedures. This makes TabAttention an alternative to the heuristic formulas that are currently used in clinical practice. We should note that while TabAttention achieved the lowest metrics among the DL models we evaluated, the differences between our approach and other DL methods using tabular data were not statistically significant, partly due to the small performance change. This small difference in the performance is likely caused by the fact that the tabular features used in TabAttention are mainly derived from the same modality (i.e. US scans), so they do not carry additional information, but instead can be considered as refined features already present in the scans. 
To develop TabAttention, we used tabular data as a hint for the network to learn attention maps and gain additional knowledge about essential aspects presented in the scans. This approach significantly improved the performance of baseline methods and demonstrated its practical applicability.

Accurate estimation of FBW is crucial in determining the appropriate delivery method, whether vaginal or Cesarean. Low birth weight (less than 2500 g) is a major risk factor for neonatal death, while macrosomia (greater than 4000 g) can lead to delivery traumas and maternal complications, such as birth canal injuries, as reported by Benacerraf \textit{et al.} \cite{benacerraf1988sonographically}. Thus, precise prediction of FBW is vital for very low and high weights. Notably, in this respect, our method is robust to outliers with high or low FBW since there is no correlation between true FBW and absolute prediction error (Pearson correlation coefficient of -0.029).

    This study has limitations. Firstly, a relatively small study cohort was used, which may affect the accuracy and generalization of the results. To address this, future work will include a larger sample size by using additional datasets. Secondly, our dataset is limited to only Caucasian women and may not be representative of other ethnicities. It is important to investigate the performance of our method with datasets from different ethnic groups and US devices to obtain more robust and generalizable results. Lastly, our method relies on fetal biometry measurements that are subject to inter- and intra-observer variabilities. This variability could potentially affect the network's performance and influence the measurements' quality. Future studies should consider strategies to reduce measurement variabilities, such as standardized protocols or automated measurements, to improve the accuracy of the method.


To summarize, we have introduced TabAttention, a new module that enables the conditional learning of attention on tabular data and can be integrated with any CNN. Our method has many potential applications, including serving as a computer-aided diagnosis tool for various clinical workflows. We have demonstrated the effectiveness of TabAttention on the FBW prediction task, utilizing both US and tabular data, and have shown that it outperforms other methods, including clinically used ones. In the future, we plan to test the method in different clinical applications where imaging and tabular data are used together. 

\section*{Acknowledgements}

This work is supported by the European Union’s Horizon 2020 research and innovation programme under grant agreement Sano No 857533 and the International Research Agendas programme of the Foundation for Polish Science, co-financed by the European Union under the European Regional Development Fund.

\bibliographystyle{splncs04}
\bibliography{bibliography}

\begin{thebibliography}{10}
\providecommand{\url}[1]{\texttt{#1}}
\providecommand{\urlprefix}{URL }
\providecommand{\doi}[1]{https://doi.org/#1}

\bibitem{bano2021autofb}
Bano, S., Dromey, B., Vasconcelos, F., Napolitano, R., David, A.L., Peebles, D.M., Stoyanov, D.: Autofb: Automating fetal biometry estimation from standard ultrasound planes. In: International Conference on Medical Image Computing and Computer-Assisted Intervention. pp. 228--238. Springer (2021)

\bibitem{benacerraf1988sonographically}
Benacerraf, B.R., Gelman, R., Frigoletto~Jr, F.D.: Sonographically estimated fetal weights: accuracy and limitation. American journal of obstetrics and gynecology  \textbf{159}(5),  1118--1121 (1988)

\bibitem{campbell1975ultrasonic}
Campbell, S., Wilkin, D.: Ultrasonic measurement of fetal abdomen circumference in the estimation of fetal weight. BJOG: An International Journal of Obstetrics \& Gynaecology  \textbf{82}(9),  689--697 (1975)

\bibitem{xgboost}
Chen, T., Guestrin, C.: {XGBoost}: A scalable tree boosting system. In: Proceedings of the 22nd ACM SIGKDD International Conference on Knowledge Discovery and Data Mining. pp. 785--794. KDD '16, ACM, New York, NY, USA (2016). \doi{10.1145/2939672.2939785}

\bibitem{duanmu2020prediction}
Duanmu, H., Huang, P.B., Brahmavar, S., Lin, S., Ren, T., Kong, J., Wang, F., Duong, T.Q.: Prediction of pathological complete response to neoadjuvant chemotherapy in breast cancer using deep learning with integrative imaging, molecular and demographic data. In: Medical Image Computing and Computer Assisted Intervention--MICCAI 2020: 23rd International Conference, Lima, Peru, October 4--8, 2020, Proceedings, Part II 23. pp. 242--252. Springer (2020)

\bibitem{guan2021predicting}
Guan, Y., Cui, H., Xu, Y., Jin, Q., Feng, T., Tu, H., Xuan, P., Li, W., Wang, L., Duh, B.L.: Predicting esophageal fistula risks using a multimodal self-attention network. In: Medical Image Computing and Computer Assisted Intervention--MICCAI 2021: 24th International Conference, Strasbourg, France, September 27--October 1, 2021, Proceedings, Part V 24. pp. 721--730. Springer (2021)

\bibitem{hadlock1985estimation}
Hadlock, F.P., Harrist, R., Sharman, R.S., Deter, R.L., Park, S.K.: Estimation of fetal weight with the use of head, body, and femur measurements—a prospective study. American journal of obstetrics and gynecology  \textbf{151}(3),  333--337 (1985)

\bibitem{holste2021end}
Holste, G., Partridge, S.C., Rahbar, H., Biswas, D., Lee, C.I., Alessio, A.M.: End-to-end learning of fused image and non-image features for improved breast cancer classification from mri. In: Proceedings of the IEEE/CVF International Conference on Computer Vision. pp. 3294--3303 (2021)

\bibitem{huang2020fusion}
Huang, S.C., Pareek, A., Seyyedi, S., Banerjee, I., Lungren, M.P.: Fusion of medical imaging and electronic health records using deep learning: a systematic review and implementation guidelines. NPJ digital medicine  \textbf{3}(1), ~136 (2020)

\bibitem{kingma}
Kingma, D.P., Ba, J.: Adam: A method for stochastic optimization. In: International Conference on Learning Representations (ICLR) (2015)

\bibitem{liu2018joint}
Liu, M., Zhang, J., Adeli, E., Shen, D.: Joint classification and regression via deep multi-task multi-channel learning for alzheimer's disease diagnosis. IEEE Transactions on Biomedical Engineering  \textbf{66}(5),  1195--1206 (2018)

\bibitem{lu2019ensemble}
Lu, Y., Zhang, X., Fu, X., Chen, F., Wong, K.K.: Ensemble machine learning for estimating fetal weight at varying gestational age. In: Proceedings of the AAAI conference on artificial intelligence. vol.~33, pp. 9522--9527 (2019)

\bibitem{scikit-learn}
Pedregosa, F., Varoquaux, G., Gramfort, A., Michel, V., Thirion, B., Grisel, O., Blondel, M., Prettenhofer, P., Weiss, R., Dubourg, V., Vanderplas, J., Passos, A., Cournapeau, D., Brucher, M., Perrot, M., Duchesnay, E.: Scikit-learn: Machine learning in {P}ython. Journal of Machine Learning Research  \textbf{12},  2825--2830 (2011)

\bibitem{plotka2022babynet}
P{\l}otka, S., Grzeszczyk, M.K., Brawura-Biskupski-Samaha, R., Gutaj, P., Lipa, M., Trzci{\'n}ski, T., Sitek, A.: Babynet: Residual transformer module for birth weight prediction on fetal ultrasound video. In: Medical Image Computing and Computer Assisted Intervention--MICCAI 2022: 25th International Conference, Singapore, September 18--22, 2022, Proceedings, Part IV. pp. 350--359. Springer (2022)

\bibitem{plotka2022deep}
P{\l}otka, S., Klasa, A., Lisowska, A., Seliga-Siwecka, J., Lipa, M., Trzci{\'n}ski, T., Sitek, A.: Deep learning fetal ultrasound video model match human observers in biometric measurements. Physics in Medicine \& Biology  \textbf{67}(4),  045013 (2022)

\bibitem{polsterl2021combining}
P{\"o}lsterl, S., Wolf, T.N., Wachinger, C.: Combining 3d image and tabular data via the dynamic affine feature map transform. In: Medical Image Computing and Computer Assisted Intervention--MICCAI 2021: 24th International Conference, Strasbourg, France, September 27--October 1, 2021, Proceedings, Part V 24. pp. 688--698. Springer (2021)

\bibitem{pressman2000prediction}
Pressman, E.K., Bienstock, J.L., Blakemore, K.J., Martin, S.A., Callan, N.A.: Prediction of birth weight by ultrasound in the third trimester. Obstetrics \& Gynecology  \textbf{95}(4),  502--506 (2000)

\bibitem{salomon2019isuog}
Salomon, L., Alfirevic, Z., Da~Silva~Costa, F., Deter, R., Figueras, F., Ghi, T.a., Glanc, P., Khalil, A., Lee, W., Napolitano, R., et~al.: Isuog practice guidelines: ultrasound assessment of fetal biometry and growth. Ultrasound in obstetrics \& gynecology  \textbf{53}(6),  715--723 (2019)

\bibitem{shaw2018self}
Shaw, P., Uszkoreit, J., Vaswani, A.: Self-attention with relative position representations. arXiv preprint arXiv:1803.02155  (2018)

\bibitem{sherman1998comparison}
Sherman, D.J., Arieli, S., Tovbin, J., Siegel, G., Caspi, E., Bukovsky, I.: A comparison of clinical and ultrasonic estimation of fetal weight. Obstetrics \& Gynecology  \textbf{91}(2),  212--217 (1998)

\bibitem{tao2021fetal}
Tao, J., Yuan, Z., Sun, L., Yu, K., Zhang, Z.: Fetal birthweight prediction with measured data by a temporal machine learning method. BMC Medical Informatics and Decision Making  \textbf{21}(1),  1--10 (2021)

\bibitem{tran2018closer}
Tran, D., Wang, H., Torresani, L., Ray, J., LeCun, Y., Paluri, M.: A closer look at spatiotemporal convolutions for action recognition. In: Proceedings of the IEEE conference on Computer Vision and Pattern Recognition. pp. 6450--6459 (2018)

\bibitem{vaswani2017attention}
Vaswani, A., Shazeer, N., Parmar, N., Uszkoreit, J., Jones, L., Gomez, A.N., Kaiser, {\L}., Polosukhin, I.: Attention is all you need. Advances in neural information processing systems  \textbf{30} (2017)

\bibitem{wang2022spatiotemporal}
Wang, X., Liu, D., Zhang, Y., Li, Y., Wu, S.: A spatiotemporal multi-stream learning framework based on attention mechanism for automatic modulation recognition. Digital Signal Processing  \textbf{130},  103703 (2022)

\bibitem{woo2018cbam}
Woo, S., Park, J., Lee, J.Y., Kweon, I.S.: Cbam: Convolutional block attention module. In: Proceedings of the European conference on computer vision (ECCV). pp. 3--19 (2018)

\bibitem{yadav2020frequency}
Yadav, S., Rai, A.: Frequency and temporal convolutional attention for text-independent speaker recognition. In: ICASSP 2020-2020 IEEE International Conference on Acoustics, Speech and Signal Processing (ICASSP). pp. 6794--6798. IEEE (2020)

\end{thebibliography}

\end{document}